\documentstyle[epsfig]{mn}

\begin{document}
\title[Q criterion modified by external field]
{Q criterion for disc stability modified by external tidal field}

\author[C.J. Jog]
       {Chanda J. Jog\thanks{E-mail : cjjog@physics.iisc.ernet.in}\\
  Department of Physics,
Indian Institute of Science, Bangalore 560012, India \\
} 


\maketitle

\begin{abstract}
The standard  $Q$ criterion (with $Q > 1$) describes the local stability of a disc supported by rotation and random motion.
 Most astrophysical discs, however, are under the influence of an external gravitational 
field which can affect their stability. A typical example is a galactic disc embedded in a dark matter halo. 
Here we do a linear perturbation analysis for a disc in an external  
field, and obtain a generalized dispersion relation and a modified stability criterion.
An external field has two effects on the disc dynamics: first, it contributes to the unperturbed rotational field, and second, it adds a tidal field term in the stability parameter. A typical disruptive tidal field results in a higher modified $Q$ value and hence leads to a more stable disc. 
We apply these results to the Milky Way, and to a low surface brightness galaxy UGC 7321. We find that
in each case the stellar disc by itself is barely stable  
and it is the dark matter halo that stabilizes the disc against local, axisymmetric gravitational instabilities. 
This result has been largely missed so far because in practice the value for $Q$ for a galactic disc 
is obtained in a hybrid fashion using the observed rotational field that is 
set by both the disc and the halo, and hence is higher than for a pure disc. 
\end{abstract}
\begin{keywords}
galaxies: kinematics and dynamics - galaxies: structure -
galaxies: internal motions - hydrodynamics - instabilities
\end{keywords}
\section{Introduction}
The usual criterion to denote stability against local, axisymmetric perturbations of an astrophysical disc supported by  differential rotation and random motion is
$$ Q = \frac {\kappa c}{\pi G \mu} > 1  \eqno (1) $$
\noindent where $\kappa$ is the local epicyclic frequency, $c$ is the one-dimensional random velocity in the radial direction or the sound speed 
in the medium, and $\mu$ is the surface density of the disc. Here $Q > 1, < 1 $ and $=1$ denote stability, instability
and marginal or neutral stability of the disc respectively (Safronov 1960, Goldreich \& Lynden-Bell 1965).
The above result is for a fluid representation of the disc, while a distribution function approach gives a value 
of 3.36 instead of $\pi$ (Toomre 1964). 
The above criterion is usually known as the Toomre $Q$ criterion.

This  criterion was obtained for an isolated disc. However, astrophysical discs are 
generally under the influence of an external field.
An obvious example is a galactic disc embedded within a dark matter halo,
or a circumnuclear disc around a black hole.
In this case the external field could play a crucial role
in deciding the stability of a disc and the above  criterion then has to be modified.
 For example, it is well-known that the dark matter halo in a galaxy
becomes progressively more important in the outer parts in terms of the total mass, yet surprisingly, its effect on the formation of instabilities in the galactic disc has not been studied too well so far. In fact, even the sign of the effect has not been identified correctly. It has often been naively assumed the dark matter halo within the disc region will 
make the disc unstable, as shown by adding the mass of halo in a certain scale height to the disc surface density in an ad-hoc way
(Corbelli 2003 , Hunter et al. 1998). Also, the inclusion of a responsive halo is said to make the disc marginally more unstable (Esquivel \& Fuchs 2007).
On the other hand, Kawata \& Hanami (1998) argued that the presence of dark matter halo makes the disc more stable, but their derivation of the effective stability criterion is ad-hoc and based on dimensional arguments.

The dark matter halo has a 3-D mass distribution while the disc has a flattened, nearly 2-D mass distribution. Further, the halo is supported by pressure while the disc is supported mainly by rotation. These differences have probably led to the confusion so far about how to incorporate the effect of the halo on the disc stability. 

In this paper we formulate the problem correctly in terms of the external field.
We show that an external field such as due to the halo has two effects on the disc dynamics: first, it contributes to the unperturbed rotational field, and second, it adds a tidal field term in the stability  parameter. 
The inclusion of a disruptive external tidal field 
 is shown to help prevent the growth of disc instabilities.  

A similar idea involving the effect of an external tidal field on a homogeneous system resulting in a modification of the standard Jeans criterion has been recently studied by Jog (2013).

Section 2 contains the derivation of the modified dispersion relation and the local stability criterion for a disc in presence of an external field. An application to the case of a galactic disc in a dark matter halo for the Galaxy and UGC 7321 is given in Section 3, and Sections 4 and 5 contain discussion and conclusions respectively.
\section {Formulation of equations}
\subsection  {Linear perturbation analysis}
We consider the local stability of a given region in an axisymmetric disc such as a galactic disc treated as a fluid.  The force equation, the continuity equation and the Poisson equation for a thin disc in an external  potential $\Phi_{ext}$ are respectively given as
$$ \frac {\partial \bf v}{\partial t} + (\bf v\: . \nabla) {\bf v} = - \frac {1}{\mu} \nabla p - \nabla \Phi - \nabla \Phi_{ext} \: , \eqno (2) $$
$$ \frac {\partial \mu}{\partial t} + \nabla . ( \mu {\bf v} ) = 0 \: , \eqno (3) $$
$$ {\nabla}^2 \Phi = 4 \pi G \mu \delta (z) \: , \eqno(4) $$
\noindent where $\delta (z)$ is the Dirac delta function, and $p=\mu c^2$ is the isothermal equation of state and $c$ is the one-dimensional r.m.s. velocity dispersion or the sound speed in the  disc. 

We use galactic cylindrical co-ordinates ($R, \phi, z$). The 
 surface density, the angular velocity and the gravitational
potential of the disc for an unperturbed state in the local region around $R$  are given respectively by $\mu_0, \Omega_0$ and $\Phi_0$.
Next, consider a perturbed disc with the perturbed quantities denoted by $\mu_1 , \Phi_1 , u, v$,  where $u, v$ are the radial and azimuthal perturbation velocity components 
respectively. Keeping only the first order terms in the perturbed quantities, and assuming the disc and the perturbation to be axisymmetric, the general linearized perturbation equations in the presence of the external field are  (e.g, Toomre 1964, 
 Binney \& Tremaine 1987) as follows
$$ \mu_0 \frac {\partial u}{\partial t}  -   2 \Omega_0  v \mu_0  +  c^2 \frac {\partial \mu_1}{\partial R} +  \mu_0 {\frac {\partial \Phi_1}{\partial R}}{\Bigr \vert}_{z=0}  +  \mu_1 \frac {\partial \Phi_{ext}}{\partial R}{\Bigr \vert}_{z=0} = 0 \: , \eqno (5) $$
$$ \frac {\partial v}{\partial t} - 2 B u = 0 \: , \eqno (6) $$
$$\frac {\partial \mu_1}{\partial t} + \mu_0 \frac {\partial u}{\partial R} = 0 \: , \eqno (7) $$
$$ \frac{1}{R}  {\Bigl [}{\frac{\partial}{\partial R}} \big ( R {\frac {\partial \Phi_1}{\partial R}}\big ) {\Bigr ]} + {\frac{{\partial}^2 \Phi_1}{{\partial}^2 z}} = 4 \pi G \mu_1 \delta (z) \: . \eqno (8)$$  
\noindent 
In linearizing the last two terms on the r.h.s. of equation (2),
the terms $\mu_0 (\partial \Phi_0 / \partial R)$ and $\mu_0 (\partial \Phi_{ext} / \partial R)$ drop out as they are taken to cancel the centrifugal force term $\Omega_0^2/ R$, 
in analogy with what is normally done for the disc-alone case (e.g, Goldreich \& Lynden-Bell 1965, Binney \& Tremaine 1987). 
  This means that the assumption of a constant undisturbed potential as in Jeans swindle for the 1-D is not required in the 2-D case (Binney \& Tremaine 1987), as discussed further.
However, the term $\mu_1 {\partial \Phi_{ext}}/ {\partial R}$ is retained, and it is this term that eventually leads to a modification  in the dispersion relation as shown next. 
Thus equation (5) is valid for a general external potential, which need not be a small perturbation on the disc potential. 

In equation (6), $B$ is the Oort constant defined to be = $(1/2) [\Omega_0 + d (\Omega_0 R)/ d R]$, so that ${\kappa}^2 = - 4 B \Omega_0$ (e.g., Binney \& Tremaine 1987).
In simplifying the continuity equation (equation 7), the term $(1/R) ({\partial}/{\partial R} [ R (\mu_0 + \mu_1) u]) $
reduces  to $(\mu_0 {\partial u}/{\partial R})$
since $u$ varies more rapidly with $R$ than does the term $(R \mu_0)$ (see Toomre 1964).

The treatment in this paper assumes the external field is not affected by the disc. This is
valid since the dark matter halo for which this calculation is applied  (Section 3) is much more massive than a galactic disc.

The treatment by Toomre (1964), and Goldreich \& Lynden-Bell (1965) was meant for the disc-alone case. However, we caution that later users have routinely used the observed or the net rotational velocity and the net $\kappa$ as arising due to the disc and the dark matter halo combined in solving these equations. 
In doing this it is implicitly assumed that
$\nabla \Phi_0$ is due to the entire galaxy (disc plus halo) that balances out the centrifugal force.
Thus the resulting $Q$ typically used for the disc in the literature is a hybrid value where the surface density is for the stellar disc and the epicyclic frequency, $\kappa$ is based on the net rotation field due to the disc and the halo. Most users of the hybrid Toomre criterion may not be aware of this inconsistency.
So by using the net $\kappa$ as done in the literature, the effect of the halo is already included in calculating the net rotational support and hence the $Q$ value. This can hide the important stabilizing effect of halo for the disc stability. This is discussed further in Section 3.

Thus the external field affects the stability of the disc in two ways. First, it affects the net $\kappa^2$ as being that due to the disc and the external field. 
The second effect is the additional factor due to the tidal field derived here. Due to the hybrid treatment in the literature as discussed above, the first effect is already taken into account by most users unwittingly. It turns out that generally the first effect is stronger when the external medium dominates the rotation of the disc as in the case of halo. 

The equations (5-7)  are linear and homogeneous, hence a linear superposition of solutions is allowed, and hence these can be solved by the method of modes. The trial solution for the perturbed quantities (surface density, and velocity components) is taken to be their respective magnitudes ${\mu_1}', u', v'$ multiplied by $exp i (wt - kr)$, where $\omega$ is the angular frequency and $k = 2 \pi/ \lambda$ is the wavenumber and $\lambda$ is the wavelength of the perturbation. The value of $k$ is chosen such that $kR >> 2 \pi$ so that the perturbation is local.

For the above trial solution, the Poisson equation (equation 8) reduces to the following (see Toomre 1964) 
$$ \frac {\partial \Phi_1} {\partial R} {\Big \vert}_{z=0}  = - i 2 \pi G \mu_1 \: .    \eqno(9) $$
For the above trial solution, and using the simplified Poisson equation (equation 9), the equations governing the perturbed motion (equations 5-7) reduce to 
$$ i w u' \mu_0 - 2 \Omega_0 v' \mu_0 + c^2 i k \mu_1 - \mu_0 i 2 \pi G \mu_1 + \mu_1 \frac {\partial \Phi_{ext}}{\partial R} {\Big \vert}_{z=0} = 0  \: , \eqno(10) $$
$$ i w v' - 2 B u' = 0 \: , \eqno (11) $$
$$ i w \mu_1 + \mu_0 i k u' = 0 \: . \eqno (12) $$
\subsection{Modified dispersion relation in external field}
On combining these and assuming  a  general non-zero perturbation, we get the following modified dispersion relation
$$ w^2 = {\kappa}^2 + k^2 c^2 - 2 \pi G k \mu_0 - \frac {{\partial}^2 \Phi_{ext}}{\partial R^2} \: , $$
$$ \: \:  \: \: \: = {\kappa}^2 + k^2 c^2 - 2 \pi G k \mu_0 + T_0  \: , \eqno (13) $$
\noindent where the last term denotes the external tidal field per unit distance along the radial direction,  calculated at the mid-plane 
at $z=0$,  defined as
$$ T_0 = \: - \:  \frac {{\partial}^2 \Phi_{ext}}{\partial R^2} \: . \eqno (14) $$
\noindent 
 Note that the first three terms on the r.h.s. of equation (13) are present in the standard dispersion relation for a disc (e.g., Jog \& Solomon 1984, Binney \& Tremaine 1987). Thus the presence of an external tidal field  adds an 
additional term T$_0$ in the dispersion relation (equation 13).
For a typical disruptive tidal field, 
 T$_0 > 0$, thus the disc will be more resistant to the onset of instabilities.

The tidal term adds to the effect due to rotation, and like the rotational term this new term also does not have a dependence on the wavenumber, $k$. Consequently, the most unstable wavelength at which $w^2$ is a minimum (corresponding to $d w^2/ d k = 0$) remain unchanged at 
$k_{min} = \pi G \mu_0 / c^2$ as shown in Section 2.2. 
Thus,  the size and the mass of the fastest growing instability remains unchanged. Normally, rotation supports the disc against the growth of large instabilities, this including the maximum wavelength of the instability that can form
will be modified in presence of the tidal field.
\subsection {Modified Q criterion in an external field}
 If $\omega^2 > 0$ for all wavenumbers $k$,
then the disc is stable against growth of perturbations. This is a second order polynomial in $k$ (see equation 13). The most unstable wavenumber, $k_{min}$ for which $\omega^2$ is a minimum, is obtained when the following is satisfied
$$ {\frac {d [\omega^2 (k)]} {d k}} = 0 \: . \eqno (15) $$
\noindent This gives a minimum since $d\omega^2 /dk^2 = c^2 > 0$ (see equation 13). Using $\omega^2$ in equation (13),
 this gives $k_{min} = \pi G \mu_0 / c^2$.
Substituting this back in equation (13), we get
$$ \omega^2 = \bigl(\frac{\pi G \mu}{c}\bigr)^2 \: \biggl[ \frac {(\kappa^2 + T_0) c^2}{(\pi G \mu)^2} - 1\biggr] \: . \eqno(16)$$ 
Define $Q_{eff}$ to be the modified or effective $Q$ parameter:
$$ Q_{eff} = \frac {({\kappa}^2 + T_0)^{1/2}  c} { \pi G \mu}   =  Q   ( 1 + T_0 / {\kappa}^2)^{1/2}  \: .   \eqno (17)   $$
\noindent Note that $Q_{eff} > 1$ denotes that the disc is stable to perturbations at all wavelengths, while $Q_{eff} < 1$ and $=1$ denotes an unstable and a marginally stable disc.
Here $Q = \kappa c/  \pi G \mu $ is the standard $Q$ criterion (equation 1). Thus, the presence of an external field adds an additional factor of $(1+T_0/\kappa^2)^{1/2}$ in the modified $Q$ parameter.
A disruptive tidal field (T$_0 > 0$) leads to a higher $Q_{eff}$, thus making the system more stable against perturbations. The inclusion of a disruptive tidal field is effectively like increasing the value of $\kappa$.  The higher the tidal field magnitude compared to $\kappa^2$, the stronger is the stabilizing effect due to the external field.

The Toomre $Q$ criterion  seems to be a surprisingly 
good indicator of star formation threshold (Kennicutt 1989), even though no account is taken of the detailed gas physics.
Strictly speaking, the $Q$ criterion is for local stability but it is often used as an indicator of the stability against non-axisymmetric perturbations. A non-axisymmetric perturbation is stabilized for $Q > 2$ (Toomre 1981, Sellwood \& Carlberg 1984).
In presence of a tidal field, $Q$ can be replaced by $Q_{eff}$ in the above criteria, thus a disruptive tidal field
makes it more difficult for star formation or spiral features to form in the regions where the dark matter halo dominates
as in the outer regions of galaxies.
\subsection {Results in an external field}

\noindent {\bf 1. Disruptive tidal field:}  The tidal field is disruptive, when  $ T_0 >  0$. A simple example of a disruptive field is that due to a point mass M$_p$, which gives T$_0$ = 2 G M$_p / R^3$ ($>0$).
In this case, even when the disc by itself is unstable to the growth of perturbations,
the addition of the last term due to the disruptive tidal field in the dispersion relation ( equation 13) makes the system more stable against the perturbations, and the effective $Q$  value is higher (see equation 17). This would be applicable for example for a circumnuclear disc  around a black hole since the latter would dominate the mass and the potential in the system. 

\noindent  {\bf 2. Compressive tidal field:} Conversely, under some conditions the tidal field could be compressive ($ T_0 <  0$), which would lower the effective $Q$ parameter, and thus tend to make the disc unstable even when it is stable by itself. A compressive tidal field seems
unphysical at a first glance, but can typically occur when the background mass distribution is smooth
and varies slowly with radius on scales larger than the perturbation (Ostriker, Spitzer \& Chevalier 1972,  Das \& Jog 1999, Renaud et al. 2009). In this case, the
gravity of the background field  enhances the gravitational field of the perturbation 
 and hence supports  the growth of perturbations rather than hindering it. 
As noted above, since $\lambda_{max}$ for the fastest growing mode remains unchanged, the mass of the instabilities also remains the same - even when the compressive tidal field supports the growth of instabilities. 

The other two tidal field components, normal to the radial direction, are always compressive (e.g., Das \& Jog 1999), but we do not consider those here. 
The tidal field is anisotropic (Das \& Jog 1999, Masi 2007), however their magnitudes are comparable for most cases, and hence their effect is felt in an average way, as shown by Jog (2013). Hence for simplicity we consider the stability of local, radial perturbations only under the radial component of the tidal field.
\section {Applications}
We apply the above results to the specific case of stability of a 
galactic disc affected by the dark matter halo it is embedded in.
 We consider two galaxies : the Milky Way as a typical spiral galaxy, and UGC 7321, a superthin, low surface brightness galaxy that is dominated by dark matter from innermost regions (Banerjee, Matthews \& Jog 2010). The aim is to compare these two cases which have different relative importance of dark matter halo.
 \subsection {Tidal field of a pseudo-isothermal halo}
We first calculate and plot the tidal field versus
radius for a dark matter halo in a galaxy and then 
discuss how this modifies the $Q$ criterion for the disc stability.
For simplicity we consider a spherical, pseudo-isothermal density distribution with the density distribution as follows
$$ {\rho}_{DM} (r) = \frac  {\rho_0 } {1 + \frac{r^2}{{R_c}^2}}  \: ,                    \eqno (18) $$
\noindent where $\rho_0$ is the central density and $R_c$ is the core radius. Such a density profile 
 is indicated by some mass models of the Galaxy (Mera et al. 1998), and also for UGC 7321 (Banerjee et al. 2010).
The corresponding potential in spherical co-ordinates is given as (see Narayan \& Jog 2002) 
$$  \Phi_{DM} (r) = - 4 \pi G \rho_0 {R_c}^2  \left [ 1 - \frac {1}{2} log (r^2 + {R_c}^2) - {\frac {R_c}{r}} 
tan^{-1} (\frac{r}{R_c}) \right] \: .                           \eqno (19)$$
Rewriting the above in cylindrical co-ordinates ($R, \phi, z$) and taking the radial derivative twice, we get the tidal field $T_0$ at the mid-plane as defined in equation (14) to be
$$ T_0 = 4 \pi G \rho_0   \left[ \frac{{R_c}^2}{R^2} - \frac{2 {R_c}^3}{R^3} tan^{-1} \frac {R} {R_c} + \frac {{R_c}^2 / R^2} {(1+ \frac {R^2}{{R_c}^2})} \right]   \: .                   \eqno(20)$$
\begin{figure}
\centering
\includegraphics[width=2.8in,height=2.4in]{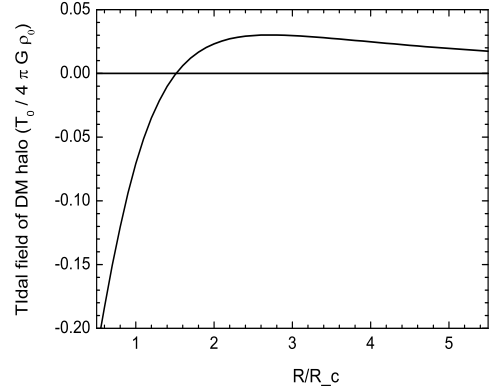}
\caption{Plot of the tidal field of a pseudo-isothermal dark matter halo T$_0/ 4 \pi G \rho_0$ in a dimensionless form, versus the galactocentric radius $R/R_C$, where R$_C$ is the core radius. The tidal field is found to be compressive in the inner region while it becomes disruptive at radii $> 1.5$ R/R$_C$.}
\end{figure}
We plot this and find that it is $>$ 0 beyond radii larger than 1.5 times the core radius, thus the tidal field due to a dark matter halo with the above density profile  is disruptive beyond this radius while it is compressive in the region inside of this (see Fig. 1).
So both compressive and disruptive field effects can be seen in this case.
\subsection {Application to the Milky Way}
We next calculate the effective $Q$ (equation 17)  for the Galactic disc embedded in the  dark matter halo with $T_0$ as obtained above for a pseudo-isothermal density profile. The halo central density, $\rho_0$ = 0.035 M$_{\odot}$ pc$^{-3}$ and the core radius, $R_C$ = 5 kpc (Mera et al. 1998). The stellar disc has an exponential surface density distribution with the central surface density $\mu_0$ = 640 $M_{\odot} pc^{-2}$ and the disc scale length, $R_D$ = 3.2 kpc (Mera et al. 1998). The stellar radial velocity dispersion $c$ is observed to be 95 km s$^{-1}$ at the centre and falls exponentially with radius with a scale length 8.7 kpc (Lewis \& Freeman 1989). 
The net $\kappa$ to be used to obtain the effective $Q$ parameter is obtained by adding the contribution from the disc and the halo in quadrature
(Section 2.1). The 
${\kappa}_{DM}^2$ due to the halo can be calculated from the potential using the standard expression $\kappa^2 =
\partial^2 \Phi/ \partial R^2 + 3 \partial \Phi/ \partial R$ obtained at the mid-plane (Binney \& Tremaine 1987). Similarly, the disc contribution is obtained 
starting from the potential for an exponential disc in the mid-plane $z=0$ (Binney \& Tremaine 1987, eq. 2.168).
\begin{figure}
\centering
\includegraphics[width=2.8in,height=2.4in]{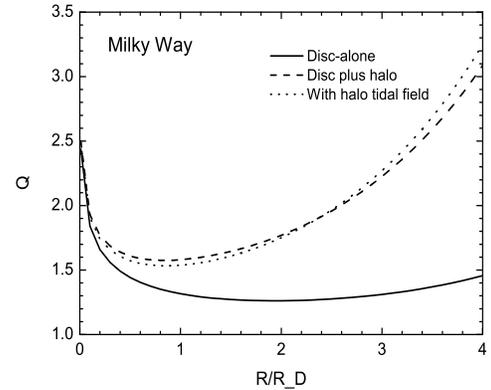}
\caption{Plot of $Q$ parameter as an indicator of local stability of the Galactic disc  vs. galactocentric radius $R/R_D$,
for disc-alone (solid line), for disc plus dark matter halo (dashed line), and $Q_{eff}$ including the  effect of the tidal field of the dark matter halo (dotted line) for the Milky Way. Note that the disc-alone is close to being unstable beyond about one disc scale length, and is stabilized due to the presence of the dark matter halo.}
\end{figure}
The resulting plots of the $Q$ parameter vs. the galactocentric radius are shown for the disc-alone case, the case with the rotation field set by disc plus halo, and the latter case including the tidal field of the halo (Fig. 2). A comparison of the curves for the
disc-alone and disc plus halo  shows that the inclusion of the halo has a striking effect in
  determining  the net
 $\kappa$ and hence the net $Q$ parameter value.
 As discussed in Section 2.1, most papers in the literature routinely use the net $\kappa$ so this effect of the halo is already taken into account in a hybrid way.
The additional effect due to the tidal field of the dark matter halo adds a factor $(1 + T_0 /\kappa^2)^{1/2}$ (equation 17), however, this makes less than 10\% difference 
in the effective $Q$ parameter (compare the dashed and the dotted curves). 

In the regions inside of 1.5 $R_C$ or about 2.5 $R_D$, the tidal filed is compressive, thus the effective $Q$ should decrease (equation 17). However,
the stellar dispersion is sufficiently high so that the net $\kappa$ still increases at lower radii. Hence though the halo tidal field is compressive 
in the inner regions, it plays little role in the determination of the $Q$ values in the central region.

In comparison, in the 1-D case, the tidal effect contributes a term that adds on to the density of the gas and can increase it by a factor of $> 10$ in centres of early-type galaxies with compressive tidal fields (Jog 2013).

Note that the values of effective $Q$ in Fig. 2 are comparable to the range of observed values (van der Kruit \& Freeman 2011), and also the range $\sim$ 1.5-2.5 that
 is seen in typical simulations of large galaxies (e.g,  Sellwood \& Carlberg 1984). In the solar neighbourhood, the value for $Q$ based on observed parameters is estimated to be $\sim 1.7$ (Binney \& Tremaine 1987). We note that this is a hybrid value since it is based on the observed rotation curve. This underscores the point made above that a hybrid $Q$ is higher than the disc-alone value (as shown in Fig. 2).
\subsection {Application to UGC 7321}
A similar plot for a low surface brightness galaxy, UGC 7321, is shown in Fig. 3. The model parameters for the halo are taken from Banerjee et al. (2010), and the
observed parameters for the stellar disc are taken from Matthews (2000), also see Banerjee \& Jog (2013). The stellar disc has a low mass with the central surface density $\mu_0$ = 50 $M_{\odot} pc^{-2}$ and the disc scale length, $R_D$ = 2.1 kpc, the stellar vertical velocity dispersion $c$ is observed to be 14.3 km s$^{-1}$ at the centre which corresponds to a central radial velocity dispersion, $c$ of 28.6 Km s$^{-1}$ assuming the velocity ratio of 2 as in the Galaxy. The dispersion is assumed to fall exponentially with radius with a scale length 2 $R_D$ = 4.2 kpc. 
The halo is dense and compact, with parameters obtained by modeling the rotation curve and the HI gas scaleheight data (Banerjee et al. 2010) as follows: the central density, $\rho_0 = 0.057$ M$_{\odot}$ pc$^{-3}$, and the core radius $R_C$ = 2.5 kpc.
Fig. 3 shows that the halo contribution to $\kappa$ and hence net $Q$ dominates from the inner regions, and could help prevent local disc instabilities.
The values of $Q$ are greater than 3 at all radii, hence the condition for prevention of non-axisymmetric instabilities (Section 2.2) is satisfied. This
could explain why spiral arms are not seen in the low surface brightness galaxies (McGaugh, Schombert \& Bothun 1995).
\begin{figure}
\centering
\includegraphics[width=2.8in,height=2.4in]{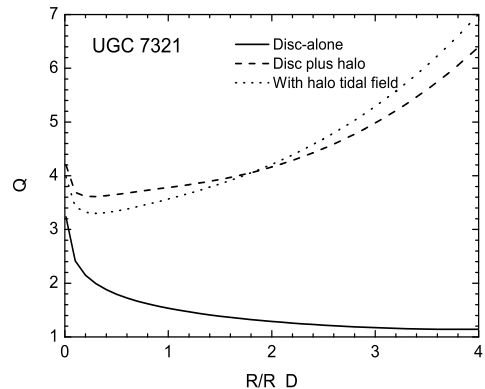}
\caption{A similar plot of $Q$ values vs. galactocentric radius $R/R_D$
for UGC 7321, a  superthin, low surface brightness galaxy. Here the dark matter halo 
dominates from the inner regions and 
results in high values of effective  $Q$ to be $ > 3$ from the inner regions. This prevents local disc instabilities, and could also help explain why such galaxies do not show non-axisymmetric spiral features.}
\end{figure}
\subsection{Stabilizing effect of halo for local disc stability}
Note that in both these galaxies, the stellar disc by itself is close to being unstable to axisymmetric perturbations ($Q \sim$ 1.2-1.3) outside one disc scalelength, and  it is the dark matter halo that stabilizes the galactic disc against the growth of local gravitational instabilities. The relative
stabilizing effect of the dark matter halo is particularly noticeable in the low surface brightness galaxy UGC 7321 which is dark-matter dominated from the inner regions. This somewhat serendipitous finding is an important result from this work.

The dark matter halo was first proposed to stabilize a galactic disc against a bar instability (Ostriker \& Peebles 1973). The halo is also known to play a crucial role in keeping the net rotation curve nearly flat at large radii (Sofue \& Rubin 2001). 
The present paper shows  that the halo is also dynamically necessary to prevent local instabilities all over the disc. 
The observed stellar velocity dispersion is too low, hence the disc is not sufficiently hot to stabilize itself against local gravitational instabilities. 

The effect of the halo on local disc stability has not been studied much, some exceptions are - Sellwood \& Carlberg (1984), Kalnajs (1987), Bottema (1993). Also, the
fact that the stars-alone disc is close to being unstable is not well-recognized, perhaps for the following reasons. First, observation of stellar velocity dispersion is difficult and was done for only a few galaxies for a long time (van der Kruit \& Freeman 1986, Bottema 1993, Herrmann \& Ciardullo 2009), though this is likely to change with the new integral field spectroscopy data (Bershady et al. 2010). Hence calculating the $Q$ values for galaxies is not easy.
Second, most modern simulations rarely consider disc-alone as this is highly unstable to a bar mode. Hence generally these studies include a halo and 
its effect on calculating the rotation field is taken into account (e.g., Bottema 2003, Fujii et al. 2011). 

Finally, we stress that the resulting hybrid value of $Q$ that is normally used is a fairly good indicator of the  stability of the disc in the halo potential since the tidal term would change this by only $< 10 \% $. However, it does not represent the stability of a true disc-alone case, and could thus underplay the effect of  halo for local disc stability.

We caution that disc stability is also affected by other physical effects. For example, inclusion of gas would tend to make the disc by itself more unstable (Jog \& Solomon 1984, Jog 1996). 
On the other hand, taking account of the finite thickness of a disc would effectively decrease the surface density and hence increase the $Q$ value (Romeo 1992). However, these effects will not qualitatively change our result that the halo is required to stabilize the disc in a typical spiral  galaxy.
\section {Discussion}
The treatment in this paper is general and could be applied to study the stability of discs in a variety of  astrophysical systems, but in this paper we focus on the application to a galactic disc in a dark matter halo.
A similar result, namely a circumnuclear disc around a central black hole being stabilized when the rotational balance by the 
central AGN (active galactic nucleus) is taken into account, was shown by Kawata et al. 2007), although they do not take account of the small tidal term effect as shown in equation (17) here.

The effect of the tidal field is likely to be important when the external field 
is not concentric with the centre of the disc.
An example 
is when two galaxies have a tidal encounter or in the extreme case when they undergo a physical collision. Here the tidal field will be destructive over most parts and have a magnitude $< \kappa^2$ for an individual galaxy. Hence the net tidal effect on the resulting $Q_{eff}$ would be small ($10- 20 \%$). But the cumulative
effect would still be significant if there are many such consecutive encounters, as say between galaxies in a group. This may explain why group galaxies do not show enhanced star formation via gravitational instabilities despite undergoing frequent gravitational encounters (Rasmussen et al. 2012). Tidal forces could also result in dragging gas out and hence
could result in a smaller burst of star formation (Di Matteo et al. 2007).

In a merging pair of galaxies, the overlapping pair would undergo a compressive field ($T_0 < 0$) and could lead to the effective $Q$ to be less than 1 (see equation 17). Hence this could be one reason for the origin of starbursts in these. This would be a distinct triggering process, in addition to the starbursts caused by shock compression of molecular clouds as proposed and studied by Jog \& Solomon (1992). The effect of tidal field would be a dynamic process, and the resulting star formation would also depend on how long the compressive fields act compared to the dynamical timescale within a disc.
 The simulations of interacting  galaxies by Renaud et al. (2009) have treated this complex problem and attributed resulting starbursts to
a compressive tidal field. The details of this physical process is likely to depend on
 the $Q_{eff}$ as proposed in this paper. An interesting point to note is that
Renaud et al. (2008) showed a correlation between the location of the young star clusters and that of the compressive tides in mergers of galaxies, this
 supports the result from the present paper.

Another case where the tidal field is non-concentric is when a galactic disc comes into the compressive core of a cluster of galaxies (Valluri 1993). We will explore some of the above cases in future work.
\section {Conclusions}
We have studied the local stability of an astrophysical disc supported by rotation and random motion, under the influence of an external
gravitational tidal field. The  modified dispersion relation contains a term denoting the tidal field, which adds to the effect due to rotation. Interestingly, the most unstable wavelength as well as the mass of the typical instability remains unchanged.
In a disruptive tidal field, the effective $Q$ value is higher, thus the disc is more stable against perturbations than by itself, whereas in contrast  a compressive tidal field helps make the disc more unstable.

 The above results are applied to a galactic disc in the tidal field of the dark matter halo. 
The halo with a pseudo-isothermal density profile is shown to be
 disruptive beyond a radius of 1.5 halo core radii (or $\sim 2.5$  disc scale lengths)  while it is compressive in the region inside of this. 

We consider two galaxies:  the Milky Way as a typical spiral galaxy, and UGC 7321, a low surface brightness galaxy that is dark-matter dominated.
We find that in both these galaxies, the stellar disc by itself is close to being unstable to local, axisymmetric perturbations, and thus it is the dark matter halo that stabilizes the galactic disc against the growth of local instabilities. 
This result is valid whether the disc is maximal as in the Galaxy (Sackett 1997) or whether the disc mass is negligible as in the low surface brightness galaxies (Bothun, Impey \& McGaugh 1997). Thus the halo is needed to ensure local disc stabilty. This point seems to be not well-appreciated in the literature, although the importance of halo in preventing a bar is well-known.   

Normally, in the literature, the value for $Q$ is obtained in a hybrid manner using the rotational field that is set both by the disc and the halo, and hence it is higher than for the disc-alone case.
While this is a fair indicator for the stability of a disc in a halo potential as in a real galaxy, it 
does not represent the stability of a true disc-alone case.
Thus, the main effect of halo on disc stability is already included in the usual hybrid value of $Q$, and the inclusion of halo tidal field can change it by at the most 10 \%.
  Hence it is unnecessasry and incorrect  to incorporate the effect of halo by adding  the halo mass to the disc mass as done for example in Corbelli (2003). 

\medskip

\noindent {\bf References}

\bigskip

\noindent Banerjee, A., Jog, C.J. 2013, MNRAS, 431, 582

\noindent  Banerjee, A., Matthews, L. D., Jog, C. J. 2010, New~A, 15, 89 

\noindent Bershady, M.A., Verheijen, M.A.W., Swaters, R.A., Andersen, D.R., Westfall, K.B., Martinsson, T. 2010, ApJ, 716, 198

\noindent Binney, J., Tremaine, S. 1987, Galactic Dynamics. Princeton Univ. Press, Princeton, New Jersey

\noindent Bothun, G.D., Impey, C., McGaugh, S.S. 1997, PASP, 109, 745

\noindent Bottema, R. 2003, MNRAS, 344, 358

\noindent Bottema, R. 1993, A\&A, 275, 16

\noindent Corbelli, E. 2003, MNRAS, 342, 199

\noindent Das, M., Jog, C.J. 1999, ApJ, 527, 600

\noindent Di Matteo, P., Combes, F., Melchior, A.-L., Semelin, B. 2007, A\&A, 468, 61

\noindent Esquivel, O, Fuchs, B. 2007, A\&A, 468, 803

\noindent Fujii, M.S., Baba, J., Saitoh, T.R., Makino, J., Kokubo, E., Wada, K. 
2011, ApJ, 730, 109

\noindent Goldreich, P., Lynden-Bell, D. 1965, MNRAS, 130, 97

\noindent Herrmann, K.A., Ciardullo, R. 2009, ApJ, 705, 1686

\noindent Hunter, D.A., Elmegreen, B.G., Baker, A.L. 1998, ApJ, 493, 595

\noindent Jog, C.J. 2013, MNRAS Letters, DOI: 10.1093/mnrasl/slt077
(also, arXiv.org/1306.4425)

\noindent Jog, C.J. 1996, MNRAS, 278, 209

\noindent Jog, C.J., Solomon, P.M. 1984, ApJ, 276, 114

\noindent Jog, C.J., \& Solomon, P.M. 1992, ApJ,  387, 152

\noindent Kalnajs, A.J. 1987, in "Dark Matter in the universe", Proceedings of IAU Symposium 117, eds. J. Kormendy \& G.R. Knapp. Reidel, Dordrecht, pg. 289 

\noindent Kawata, D., Cen, R.,  Ho, L.C. 2007, ApJ, 669, 232

\noindent Kawata, D.,  Hanami, H. 1998, PASJ, 50, 547

\noindent Kennicutt, R.C. 1989, ApJ, 344, 685

\noindent Lewis, J.R.,  Freeman, K. C. 1989, AJ, 97, 139 

\noindent Masi, M. 2007,  Am. J. Ph., 75,  116

\noindent Matthews, L.D. 2000, AJ, 120, 1764

\noindent McGaugh, S.S., Schombert, J.M., Bothun, G.D. 1995, AJ, 109, 2019

\noindent Mera, D., Chabrier, G., Schaeffer, R. 1998, A\&A, 330, 953

\noindent Narayan, C.A.,  Jog, C.J. 2002, A\&A, 394, 89

\noindent Ostriker, J.P., Peebles, P.J.E. 1973, ApJ, 186, 467

\noindent Ostriker, J.P., Spitzer, L. Chevalier, R.A. 1972, ApJ, 176, L51

\noindent Rasmussen, J., Mulchaey, J.S., Bai, L., Ponman, T.J., Raychaudhury, S., Dariush, A. 2012, ApJ, 757, 122

\noindent Renaud, F., Boily, C.M., Fleck, J.-J., Naab, T., Theis, Ch. 2008, MNRAS, 391, L98

\noindent Renaud, F., Boily, C. M., Naab, T., Theis, Ch.
2009, ApJ, 706, 67

\noindent Romeo, A.B. 1992, MNRAS, 256, 307

\noindent Sackett, P. D. 1997, ApJ,  483, 103

\noindent Safronov, V. S. 1960, Ann d'Ap, 23, 979

\noindent  Sellwood, J.A., Carlberg, R.G. 1984, ApJ, 282, 61

\noindent Sofue, Y., \& Rubin, V. 2001, ARAA, 39, 137

\noindent Toomre, A. 1981, in Structure and dynamics of normal galaxies, ed. S.M. Fall\& D. Lynden-Bell. Cambridge Univ. Press, Cambridge, pg. 111

\noindent Toomre, A. 1964, ApJ, 139, 1217

\noindent Valluri, M. 1993, ApJ, 408, 57

\noindent van der Kruit, P.C., Freeman, K.C. 2011, ARAA, 49, 301

\noindent van der Kruit, P.C., Freeman, K.C. 1986, ApJ, 303, 556

\end{document}